\documentclass[prb,amsmath,amssymb]{revtex4}
\usepackage{graphicx}
\usepackage{bm}

\begin{document}


\title{Closed, spirograph-like orbits in power law central potentials}

\author{M. A. Reynolds}
\email{anthony.reynolds@erau.edu}
\author{M. T. Shouppe}
\affiliation{Department of Physical Sciences,
Embry-Riddle Aeronautical University,
Daytona Beach, Florida, 32114}

\date{\today}

\begin{abstract}
Bertrand's theorem proves that inverse square and Hooke's law-type
central forces are the only ones for which all bounded orbits are closed.
Similar analysis was used to show that for other central force laws there exist closed orbits
for a discrete set of angular momentum and energy values.
These orbits can in general be characterized as ``spirograph''-like,
although specific orbits look more ``star''-like or ``triangular.''
We use the results of a perturbative version of Bertrand's theorem to predict which values of angular momentum
and energy result in closed orbits,
and what their shapes will be.

\vspace{1cm}
This article has been submitted to the American Journal of Physics.
After it is published, it will be found at \verb"http://scitation.aip.org/ajp/".
\end{abstract}

\pacs{Valid PACS appear here}
\maketitle

\newcommand{\Torb}{T_{orb}}
\newcommand{\Trad}{T_{rad}}
\newcommand{\eff}{\textit{\scriptsize{eff}}}

\section{Introduction}

It is well known that there are only two central force laws for
which all bounded orbits are closed.\cite{Bertrand,Goldstein80}
By ``closed,'' we mean that the orbiting object returns to the same spatial location with the same velocity
in a finite amount of time
(specifically, it returns to the same location in phase space).
By ``bounded,'' we mean that the distance $r$ between the orbiting object
and the central object always remains between two fixed values, $r_{min} \le r \le r_{max}$,
called the radial turning points,
or in the case of elliptical orbits they are called periapsis and apoapsis.
This result is known as Bertrand's theorem,\cite{Bertrand}
first obtained in 1873.

If the attractive force is represented by a power law, $F = -kr^\lambda$,
then only $\lambda=-2$ (an inverse square force given by Newton's law of gravitation or Coulomb's law)
and $\lambda=1$ (a spring-like force given by Hooke's law) admit closed orbits,
both of which happen to be elliptical.
In fact,
the orbits in these two potentials satisfy the additional criterion that they are ``non-crossing.''
A bounded orbit does not cross itself in configuration space
if the ratio of its orbital period, $\Torb$,
to the period of its radial oscillations, $\Trad$, is an integer.
Here,
we use the parameter $\beta \equiv \Torb/\Trad$ to denote this ratio.
In the case of Newtonian gravity $\beta = 1$,
which means that there is only one periapsis and one apoapsis per orbit,
and the central body resides at one focus of the elliptical orbit.
In the case of Hooke's law, however, $\beta=2$,
and the central body is located at the center of the elliptical orbit.
This means that there are \textit{four} turning points (two close, two far)
in each orbit.

The implications of
Bertrand's theorem have been investigated extensively,
ranging
from the symmetries inherent in the potentials
to the deep connections between classical and quantum mechanics that it reveals.\cite{Grant94}
The fact that an orbit is closed means that, besides energy and angular momentum,
there must be an additional conserved quantity --- the Runge-Lenz vector.\cite{ref:RungeLenz}
Also, closely related to the fact that only $\lambda = -2$ and $\lambda = 1$ admit closed classical orbits
is the result that these two potentials result in an exactly solvable Schrodinger equation.\cite{Wu00}
In addition, these two potentials are ``dual'' in the sense that
one problem can be obtained from the other by a change of variable.\cite{Grant94}
In addition, many authors have obtained proofs of the theorem that are more elegant and pedagogical
than the original,\cite{Brown78,Zarmi02,Grandati08}
and central potentials other than power-law have been investigated.\cite{Rodriguez98}

In this work we focus on analytical methods suitable for the intermediate mechanics student,
as well as numerical techniques that can be used to find closed orbits (especially those with interesting shapes)
in central forces other than inverse square or Hooke's law.
The types of closed orbits that can be obtained are introduced in Sec.~\ref{sec:classify}.
In Sec.~\ref{sec:BrownBertrand},
Bertrand's proof of his eponymous theorem is briefly outlined,
and a more pedagogical proof, first given by Brown,\cite{Brown78}
is covered in detail.
This detail is needed
because Brown's method includes the mathematical insight necessary to analyze large amplitude perturbations from stable
circular orbits.
Finally,
in Sec.~\ref{sec:finitepert} we obtain conditions that must be satisfied so that these large-amplitude orbits are closed,
and several representative trajectories are obtained numerically.

\section{Closed orbit classification}
\label{sec:classify}

For all power law central forces other than inverse-square and Hooke's law,
most orbits, while they may remain bounded, are not closed.
However,
there are three cases in which the orbits \textit{are} closed.
First,
as long as $\lambda > -3$,
all power law central forces exhibit a stable, closed, circular orbit at the
radial location $r_0$ where the effective force $F_\eff$ is zero
\begin{equation}\label{eq:Feff}
F_\eff(r_0) = -kr_0^\lambda + \frac{L^2}{mr_0^3} = 0 ,
\end{equation}
where the second term is the ``centrifugal force,''
$L$ is the (constant) angular momentum,
and $m$ is the mass of the orbiting object.
This stable radial location is given by $r_0^{\lambda+3} = L^2/mk$.
Of course,
if $\lambda \le -3$ there are no stable circular orbits.
Indeed, when $\lambda = -3$, Newton showed that the trajectory is a
so-called Cotes' spiral.\cite{Danby}

Any deviation from a circular trajectory allows the possibility that
the orbit may no longer be closed.
However,
the second case in which closed orbits arise is
when the orbiting object is perturbed only slightly from the stable circular orbit.
If $r$ is infinitesimally
close to $r_0$, then the effective potential energy
(defined as usual by $F_\eff = - dU_\eff/dr$)
can be expanded about $r_0$
\begin{equation}
\label{eq:Ueff}
U_\eff \approx U(r_0) + \frac{1}{2} k_\eff (r-r_0)^2 ,
\end{equation}
where the effective spring constant is
\begin{equation}
k_\eff = (\lambda + 3) \frac{L^2}{mr_0^4} .
\end{equation}
Two types of periodic motion are now superposed,
the previous orbital motion as well as a radial oscillation in the simple-harmonic
effective potential of Eq.~(\ref{eq:Ueff}).
The period of these radial oscillations is $\Trad = 2\pi \sqrt{m/k_\eff}$.
Since the orbital period, obtained by taking a ratio of the circumference, $2\pi r_0$,
to the orbital velocity, $\sqrt{kr_0^{\lambda+1}/m}$,
is given by $\Torb = 2\pi\sqrt{mr_0^{1-\lambda}/k}$,
the ratio of the two periods is\cite{ref:smallperturb}
\begin{equation}\label{eq:betalambdaplus3}
   \beta \equiv \frac{\Torb}{\Trad} = \sqrt{\lambda+3} .
\end{equation}
If $\lambda$ is such that $\Torb/\Trad$ is a rational fraction, $\beta = p/q$,
where $p$ and $q$ are integers,
then this ``almost-circular'' orbit will be closed.
Hence, for certain forms of the power law that satisfy
\begin{equation}
\label{eq:lambdasmall}
   \lambda = \frac{p^2}{q^2} - 3 ,
\end{equation}
orbits that are only slightly perturbed from a circular orbit are closed.
Bertrand\cite{Bertrand} was able to show that in two special cases, $p=q$ and
$p=2q$, corresponding to $\lambda = -2$ and $\lambda = 1$, respectively,
orbits with large (not just infinitesimal) deviations from a circular trajectory remain closed.

This analysis suggests that other solutions of Eq.~(\ref{eq:lambdasmall}),
e.g., $\lambda = 6$ ($p=3q$)  and $\lambda = 13$ ($p=4q$),
admit closed orbits for infinitesimal perturbations from a circular orbit.
However,
we show in Sec.~\ref{sec:finitepert} that for \textit{all} values of $\lambda$
that exhibit stable circular orbits,
\textit{finite} perturbations from a circular orbit can result in
values of $\beta$ that are rational fractions.
This is the third case, mentioned above, in which the orbits are closed.\
In this situation,
most of the
energy-angular momentum parameter space
results in non closed orbits,
but certain discrete values of these two parameters result in closed orbits.
Further, most of these are ``crossing'' orbits in which the trajectory crosses itself one or more times
before returning to the original location,
which means that they correspond to a rational fraction where $q \ne 1$.
There are a few instances in which $\beta$ is an integer, though,
and these orbits can be triangular ($\beta=3$) or even square ($\beta=4$)
in shape.
These large values of $\beta$, however,
require large positive values of $\lambda$.

\subsection{Central forces that are not power laws}

Of course,
the previous conclusions, as well as the analysis below,
are not limited to power law central forces.
Gauss's Law implies that
an arbitrary (but spherically symmetric) mass density distribution $\rho(r)$ results
in a central force law
\begin{equation}
F_r \sim - \frac{1}{r^2} \int_0^r \rho(r') r'^2 dr' .
\end{equation}
For any particular density distribution of interest,
the effective force and potential energy,
the stable circular orbit radius,
and the ratio of the orbital and radial oscillation periods,
Eqs.~(\ref{eq:Feff})-(\ref{eq:betalambdaplus3}),
can all be obtained.
Any parameters describing $\rho(r)$ will of course replace $\lambda$.
For power law central forces,
the self-consistent density distribution is
\begin{equation}
\rho(r) \sim (\lambda + 2) r^{\lambda -1} .
\end{equation}
In fact, quite a bit of theoretical work has been done on the problem of orbits
in the gravitational potentials of galaxies and globular clusters.\cite{Spitzer87,BinneyTremaine}
For example,
Adams and Bloch\cite{Adams05} analyzed orbits in the
so-called Hernquist potential
\begin{equation}
   F_r \sim - \frac{1}{(r+r_s)^2} ,
\end{equation}
where $r_s$ is the length scale of the potential,
and the potential is due to an extended mass distribution with density
$\rho(r) \sim 1/r(r+r_s)^3$.
This distribution turns out to be a good approximation for elliptical galaxies
and dark matter haloes.
The focus in these studies has been on understanding how the orbits affect the dynamics of the system,
and not on whether each individual orbit is closed or not.
Also, Struck\cite{Struck06} was able to analytically solve for the orbits using the so-called ``epicycloid'' approximation,
which assumes the orbit is a precessing ellipse
whose shape can be expressed as a function of the type
\begin{equation}
   \frac{1}{r(\theta)} = f(\theta) \left[ 1 + e \cos(1-b)\theta \right]
\end{equation}
where $e$ is the eccentricity and $b$ determines the precession rate.
Of course,
the parameters $e$ and $b$, along with the function $f(\theta)$ must be determined from the form
of the potential.
This technique allowed him to obtain the result in Eq.~(\ref{eq:betalambdaplus3}) above,
and therefore obtain orbital resonance conditions that can assist understanding galactic dynamics,
such as bars in spiral galaxies.

\section{Finite radial perturbations}
\label{sec:BrownBertrand}

Bertrand\cite{Bertrand} used the well-known orbit equations\cite{Goldstein35} to express
$\Delta\theta$ as an integral over the radial excursion $r$,
where $\Delta\theta$ is the angle swept out by the trajectory.
In order for the orbit to be closed,
he then required that this integral,
when evaluated between two neighboring turning points,
be a rational fraction times $\pi$, or in our notation,
\begin{equation}\label{eq:thetar2}
   \Delta\theta = \frac{\pi}{\beta} = \int_{r_{min}}^{r_{max}} \frac{dr}{r^2 \sqrt{\frac{2m}{L^2}[E-V(r)] - \frac{1}{r^2}}} ,
\end{equation}
where $r_{min}$ and $r_{max}$ are roots of the denominator.
He took a global approach,
simultaneously expanding the integral for small oscillations about a stable circular orbit
as well as letting $r_{max}\rightarrow \infty$.
He was then able to show that the requirement in Eq.~(\ref{eq:thetar2})
means that $V(r)$ must be a power law with $\lambda=-2$ or $\lambda= 1$.
Unfortunately, his proof does not easily divulge any physical insight.
On the other hand,
Brown's method,\cite{Brown78}
in which he solved for the periodic motion in the anharmonic potential [see Eq.~(\ref{eq:Ueffcubic})]
near the radius of the stable circular orbit,
not only proves Bertrand's theorem,
but also allows the derivation of a closed orbit criterion that is valid for any power $\lambda$.

Here we outline Brown's method, and quote the results that are relevant to the present discussion.
First, he solved the dynamical equation for radial motion in the potential given by Eq.~(\ref{eq:Ueff})
by assuming that the object is in an initially stable, circular orbit with $r=r_0$
and orbital speed $v_{\theta,0} = \sqrt{kr_0^{\lambda + 1}/m}$.
Then a small radial impulse is imparted to the object (in order to conserve the angular momentum $L$)
which results in a nonzero radial velocity $v_{r,0}$.
Of course, the subsequent trajectory consists of a harmonic oscillation of the radial coordinate, $r$,
\begin{equation}\label{eq:rone}
r(t) = r_0 + \epsilon \cos\omega_0 t ,
\end{equation}
where $\omega_0^2 = k_\eff/m$ is just the frequency of small radial oscillations, as we obtained above,
and $\epsilon$ is the amplitude of the radial oscillations.
There is a simple relation between the initial radial velocity $v_{r,0}$ and the amplitude $\epsilon$,
which is $v_{r,0} = \omega_0 \epsilon$, or
\begin{equation}\label{eq:vr0andepsilon}
\frac{v_{r,0}}{v_{\theta,0}} = \sqrt{\lambda + 3} \left( \frac{\epsilon}{r_0} \right) ,
\end{equation}
and which comes from elementary simple-harmonic-motion theory.\cite{energy}

Bertrand's theorem, however,
is a statement about the character of \textit{finite} radial oscillations,
and the restriction to infinitesimal amplitudes must therefore be relaxed.
It turns out that it is sufficient to retain one more term,
the cubic term, in the expansion in Eq.~(\ref{eq:Ueff}),
which becomes
\begin{equation}
\label{eq:Ueffcubic}
U_\eff \approx U(r_0) + \frac{(\lambda + 3)}{2} \, \frac{L^2}{mr_0^4} (r-r_0)^2 +
\frac{(\lambda^2 - \lambda - 12)}{6} \, \frac{L^2}{mr_0^5} \, (r-r_0)^3 ,
\end{equation}
and then apply the classic solution to this anharmonic oscillator problem, which was given by Landau and Lifshitz.\cite{Landau76}
The technique consists of seeking a solution that is a series of ``successive approximations.''
The first order approximation is just Eq.~(\ref{eq:rone}),
while the second and third order approximations include oscillations at harmonics of the fundamental frequency,
$\cos 2\omega t$ and $\cos 3\omega t$.
Here,
$\omega$ is the exact anharmonic oscillation frequency,
slightly shifted from $\omega_0$ by a term that is proportional to $\epsilon^2$
\begin{equation}
\omega^2 = \omega_0^2 \left[ 1 - \frac{(\lambda - 10)(\lambda - 1)}{12} \left( \frac{\epsilon}{r_0} \right)^2 \right] .
\end{equation}
A well-known example of this effect is
the large-amplitude pendulum,
whose exact restoring force is proportional to $\sin\theta \approx \theta - \theta^3/3!$,
and an inclusion of the cubic term results in
an amplitude-dependent period.

Brown also showed that for large amplitudes
the angular velocity of the orbital motion is also slightly shifted by a term that is proportional
to $\epsilon^2$.
Using his notation
\begin{equation}
\langle \dot{\theta} \rangle^2 = \langle \dot{\theta} \rangle_0^2
   \left[ 1 + (\lambda - 1) \left( \frac{\epsilon}{r_0} \right)^2 \right] ,
\end{equation}
where $\dot{\theta}$ is the angular orbital velocity,
the brackets $\langle \rangle$ indicate an average over one orbital period,
and the subscript $0$ denotes the stable circular orbit value in the limit $\epsilon \rightarrow 0$.
Retaining only terms of lowest order in $\epsilon$,
the ratio of the two periods is
\begin{equation}\label{eq:Brownresult}
  \beta^2 =  \frac{\Torb^2}{\Trad^2} = \frac{\omega^2}{\langle \dot{\theta} \rangle^2}
          \approx  (\lambda+3)
          \left[ 1 - \frac{(\lambda - 1)(\lambda + 2)}{12} \left( \frac{\epsilon}{r_0} \right)^2 \right] .
\end{equation}
This is Brown's main result, and it proves Bertrand's theorem.
For \textit{all} orbits to be closed,
the ratio of the two periods, $\beta$, must be \textit{independent} of the radial amplitude,
and this is only true when the coefficient of $\epsilon^2$ is zero.
That is, $\lambda=1$ or $\lambda= -2$, as previously stated.
As it must,
Eq.~(\ref{eq:Brownresult}) also contains the limit given in Eq.~(\ref{eq:betalambdaplus3}),
which might be called a ``restricted version'' of Bertrand's theorem:
    ``For infinitesimal perturbations, $\epsilon \rightarrow 0$,
    Eq.~(\ref{eq:Brownresult}) reduces to
    Eq.~(\ref{eq:betalambdaplus3}),
    which means that the condition for closed orbits is
    Eq.~(\ref{eq:lambdasmall}).''

\begin{figure}
\includegraphics[width=4in]{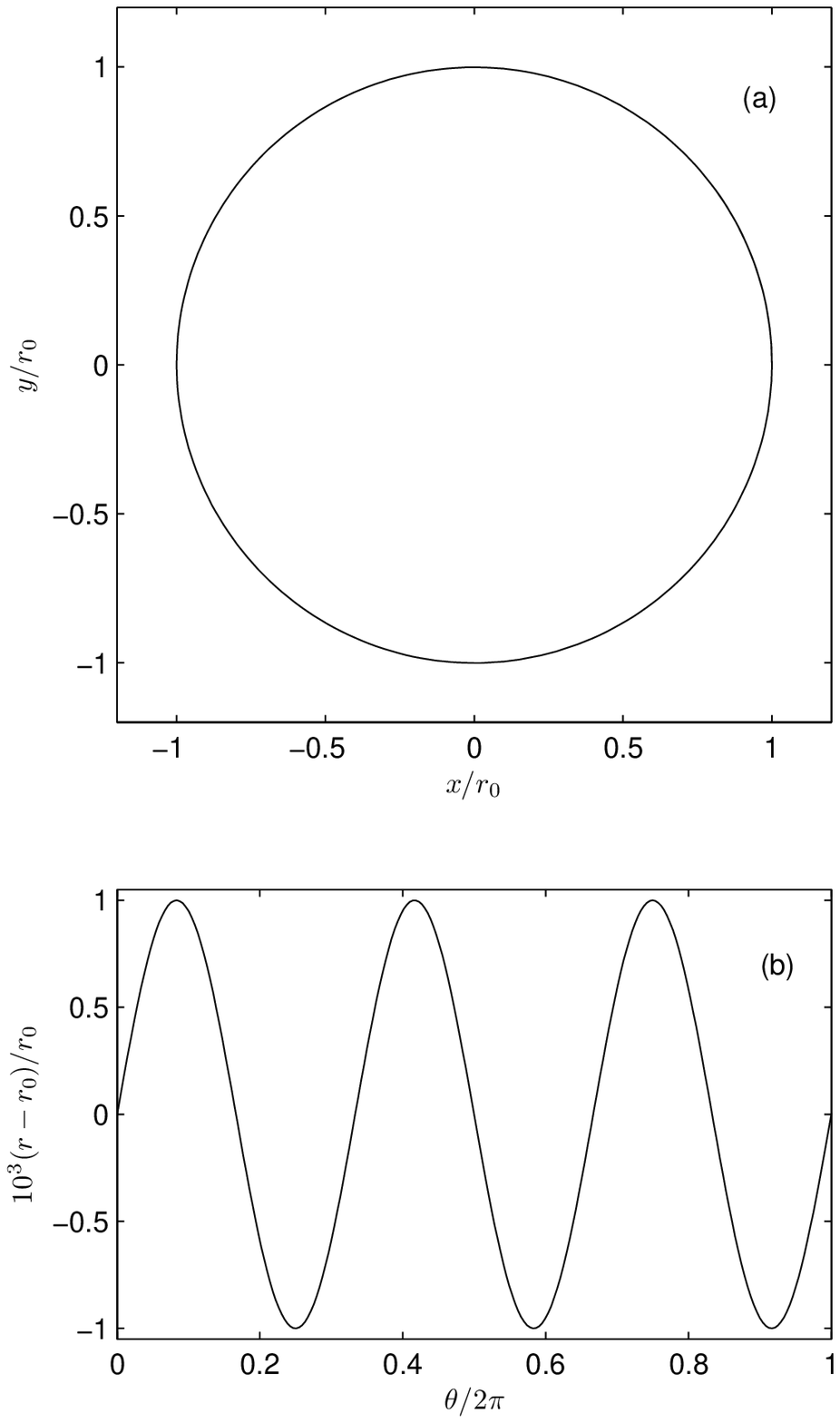}
  \caption{(a) Trajectory in the orbital plane ($x$, $y$) for the case $\lambda = 6$,
  with the initial conditions $x_0 = r_0$, $y_0 = 0$,
  and $v_{r,0} / v_{\theta,0} = 0.003$.
  (b) Radial location, $r=\sqrt{x^2+y^2}$, as a function of angular position $\theta = \arctan(y/x)$.
  The three radial excursions of the first orbit can be clearly seen,
  with an amplitude $\epsilon/r_0 = 10^{-3}$,
  as well as the fact that when $\theta = 2\pi$, $r$ returns to $r_0$.}
  \label{fig:lambda61}
\end{figure}

\begin{figure}
\includegraphics[width=4in]{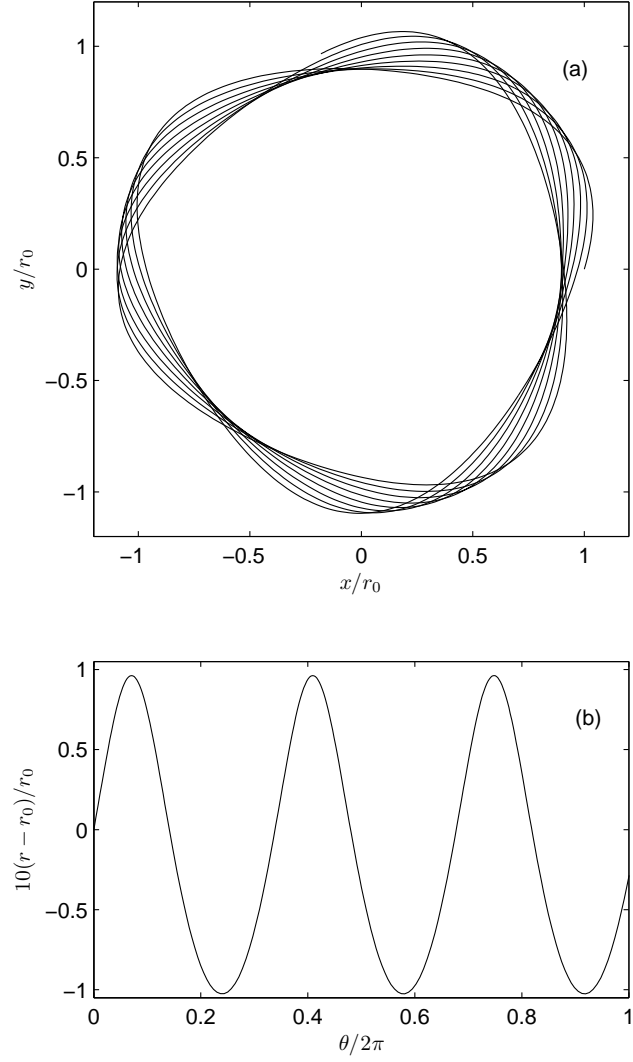}
  \caption{Same as Fig.~\ref{fig:lambda61}, except that $v_{r,0} / v_{\theta,0} \equiv 0.3$.
  This implies that $\epsilon/r_0 \approx 0.1$, and $\beta \approx 2.95$.
  This means that the period of radial oscillations is slightly more than one third of the orbital period,
  so that when $\theta = 2\pi$,
  the radial oscillation should have a phase $2\pi \times 2.95$,
  and a displacement of $0.1 \sin (2\pi \times 2.95) = -0.032$,
  as observed.}
  \label{fig:lambda62}
\end{figure}

From a practical perspective, however,
to integrate Newton's second law numerically and obtain a trajectory,
it is the initial conditions, $v_{r,0}$ and $v_{\theta,0}$,
that must be specified.
In addition, it is the parameter $\beta$ that is of primary interest,
not the radial amplitude $\epsilon$.
It is useful, therefore,
to eliminate $\epsilon$ from Eq.~(\ref{eq:Brownresult}), using Eq.~(\ref{eq:vr0andepsilon}), to obtain
\begin{equation}\label{eq:Brownresultv}
  \beta^2 \approx  (\lambda+3)
           - \frac{(\lambda - 1)(\lambda + 2)}{12} \left( \frac{v_{r,0}}{v_{\theta,0}} \right)^2 .
\end{equation}
Trajectories that demonstrate the restricted version of Bertrand's theorem
(obtained by numerically integrating Newton's second law using a Runge-Kutta 4th order method)
are shown in
Figs.~\ref{fig:lambda61} and \ref{fig:lambda62}
for a force law parameter $\lambda=6$.\cite{lambda6}
In Fig.~\ref{fig:lambda61},
an initial condition of $v_{r,0} / v_{\theta,0} = 0.003$
results in an almost circular orbit.
Since the initial radial velocity is small, the radial amplitude is likewise small,
and Eq.~(\ref{eq:vr0andepsilon}) predicts $\epsilon/r_0 \approx 10^{-3}$ for the parameters chosen,
which agrees with the numerical result shown in Fig.~\ref{fig:lambda61}(b).
In addition,
Eq.~(\ref{eq:betalambdaplus3}) predicts $\beta = 3$,
which is also seen clearly in Fig.~\ref{fig:lambda61}(b),
even though the radial oscillation is not perceptible in Fig.~\ref{fig:lambda61}(a).
All trajectories in this paper share the following initial conditions:
$x_0 = r_0$,
$y_0 = 0$,
and
$v_{y,0} = v_{\theta,0} = \sqrt{kr_0^{\lambda + 1}/m}$.
This means that if $v_{r,0}=0$, then the orbit is stable and circular.
It also means that the angular momentum remains fixed.
Varying the initial radial velocity changes the orbit shape because the total energy varies.

When the radial impulse imparts a large radial velocity, say $v_{r,0} / v_{\theta,0} = 0.3$,
the closed nature of the orbit is lost,
even though it is still bounded.
This can be seen in Fig.~\ref{fig:lambda62}.
For $\lambda = 6$,
Eq.~(\ref{eq:Brownresult}) becomes
\begin{equation}\label{eq:Brownresult6}
  \beta^2 \approx  9
           - \frac{10}{3} \left( \frac{v_{r,0}}{v_{\theta,0}} \right)^2 ,
\end{equation}
or
$\beta \approx  3 - (5/9) \left( v_{r,0} / v_{\theta,0} \right)^2$,
which shows that the orbital period decreases to \textit{less} than three times the radial oscillation
period as the radial amplitude increases.
This is indicated in Fig.~\ref{fig:lambda62}(b) by the fact that the radial position does not quite
return to $r_0$ after one complete orbit.
We can confirm this mismatch quantitatively
using Eq.~(\ref{eq:Brownresult6}), which gives
$\beta \approx 2.95$,
and this means that
when $\theta = 2\pi$,
the radial oscillation should have a phase $2\pi \times 2.95$,
and a displacement of $0.1 \sin (2\pi \times 2.95) = -0.032$,
and this is just what is observed in Fig.~\ref{fig:lambda62}(b).
The amplitude is also consistent,
for Eq.~(\ref{eq:vr0andepsilon}) predicts
$\epsilon/r_0 \approx 0.1$,
which again agrees with the numerical result in Fig.~\ref{fig:lambda62}(b).

\section{Conditions for closed orbits}
\label{sec:finitepert}

Now that we have confirmed numerically the restricted version of Bertrand's theorem,
along with the fact that the orbit does not remain closed when the radial amplitude is not infinitesimal,
we can now investigate the conditions that allow large amplitude orbits (in power laws other than $\lambda = -1, \; 2$)
to be closed.
In fact, Eq.~(\ref{eq:Brownresultv}) is just such a condition.
Above,
we used
Eq.~(\ref{eq:Brownresultv}) to predict the value of $\beta$
(and whether it is a rational fraction or not)
from a knowledge of the initial conditions (e.g., $v_{r,0}$)
and it worked
as long as $v_{r,0}$ was small.
Now, however, it is clear that
Eq.~(\ref{eq:Brownresultv}) also indicates that there can be closed orbits for \textit{any} value of $\lambda$,
as long as $v_{r,0}$ has the correct value.
To see this,
invert
Eq.~(\ref{eq:Brownresultv}) to obtain $v_{r,0}$ as a function of $\beta$
\begin{equation}\label{eq:Browninvert}
  \left( \frac{v_{r,0}}{v_{\theta,0}} \right)^2 \approx
  \frac{12}{(\lambda - 1)(\lambda + 2)}
          \left( \lambda + 3 - \beta^2 \right)  .
\end{equation}
In this case,
we first choose the force law parameter $\lambda$
and then the desired ratio of the periods, $\beta$.
Then, Eq.~(\ref{eq:Browninvert}) predicts the initial radial velocity needed to obtain that
particular closed orbit.
Of course,
the larger that the difference is between $\beta$ and $\sqrt{\lambda + 3}$,
the larger the radial oscillation,
and Eq.~(\ref{eq:Browninvert}) represents a poorer approximation.

For example,
again considering the force law parameter $\lambda = 6$,
Eq.~(\ref{eq:Browninvert}) reduces to
\begin{equation}\label{eq:Browninvert6}
  \left( \frac{v_{r,0}}{v_{\theta,0}} \right)^2 \approx
  \frac{3(9 - \beta^2)}{10} .
\end{equation}
It is clear that $\beta=3$ is the small radial oscillation limit since it predicts
an initial radial velocity of $v_{r,0} = 0$.
In addition,
since $v_{r,0}$ must be real,
$\beta$ will always be \textit{less than} 3 as the orbit deviates from a stable circle.
This fact was already clear from Eq.~(\ref{eq:Brownresult}).
As $v_{r,0}$ increases from zero,
$\beta$ will take on a continuum of real values less than 3,
most of which will not be rational.
However,
$\beta$ \textit{will} pass through an infinite number of discrete values that are rational,
implying that the corresponding orbit will be closed.
For the trajectory in Fig.~\ref{fig:lambda62}, $\beta \approx 2.95$,
and it is probably not rational,
since it was obtained by fixing $v_{r,0}$.
A rational fraction near this value is $\beta = \frac{295}{100} = \frac{59}{20}$,
which means that the orbit will have 59 radial oscillations for every 20 orbits about the center.
Such an orbit is shown in Fig.~\ref{fig:lambda63},
which is clearly closed with the correct value of $\beta$.
However,
the initial radial velocity needed to obtain this orbit is not quite the prediction of Eq.~(\ref{eq:Browninvert6}),
which is $v_{r,0} / v_{\theta,0} \approx 0.29875$.
This is because although Eq.~(\ref{eq:Browninvert6}) follows from Eq.~(\ref{eq:Brownresult}),
which is valid for large enough radial amplitudes to prove Bertrand's theorem,
it represents a poorer approximation as $(3-\beta)$ increases.
In order to determine the correct value of $v_{r,0}$ needed for such an orbit,
a more sophisticated numerical technique is required.

\subsection{Numerical determination of closed orbits}

There are two methods that can be used to find the necessary value of $v_{r,0}$ that results in an orbit with a particular $\beta$:
brute force search and root finding.
Both methods can successfully utilize the technique of Poincar\'{e}'s surface of section,\cite{Berge84}
which takes the continuous time evolution of a high-dimensional trajectory
and replaces it with a discrete mapping in fewer dimensions, usually two.
In the present case,
we plot in radial phase space (i.e., $\dot{r}$ versus $r$) the locations where a particular
trajectory crosses the positive $x$ axis,
for example.
Then,
closed orbits can be found when the trajectory returns to the same phase space location after an integral number of orbits.
The surface of section for the trajectory in Fig.~\ref{fig:lambda63} is shown in Fig.~\ref{fig:lambda6phase}.
Since the initial conditions were $r=r_0$ and $\dot{r}=v_{r,0}$, where $v_{r,0}$ is positive,
the initial location in Fig.~\ref{fig:lambda6phase} is denoted by a circle.
After 20 orbits,
and therefore 20 crossings of the positive $x$ axis,
the trajectory returns to the same phase space location.
This confirms that the orbit is closed.
In fact,
the trajectory can be followed for several ``recurrence periods,''
i.e., 40 or 60 orbits,
to make sure that the Poincar\'{e} section is periodic in the long term.

In addition to confirming that the orbit is closed,
the surface of section suggests a technique that works for the second method:
root finding \textit{via} the shooting method.\cite{NumericalRecipesShooting}
Here, the shooting method works in the standard way,
by casting the problem as a two-point boundary value problem.
The initial condition is varied --- in this case $v_{r,0}$ (the other three initial conditions,
$x_0$, $y_0$, and $v_{\theta,0}$, remain fixed) --- and the equation of motion
is integrated until the desired final condition is obtained.
The final condition here is that for an orbit with $\beta = \frac{p}{q}$,
the distance in phase space between the ``zeroth'' crossing of the
positive $x$ axis and the $q$th crossing be zero,
i.e., they must be identical.
Of course, a good initial guess for $v_{r,0}$ is needed,
and this is supplied by Eq.~(\ref{eq:Browninvert}).
In addition,
a robust root-finding method must be employed.
Since the derivative of our ``function'' (distance in phase space as a function of $v_{r,0}$) is not available analytically,
and since the tolerance of the root-finding method should not exceed the tolerance of the numerical integration,
the simple secant method should work fine.
On the other hand,
since the phase space distance is a positive definite quantity,
the desired distance is not just a root, but also a minimum.
For this reason,
a minimization method, such as Brent's method,\cite{NumericalRecipesBrent} can also be used.
It turns out that in practice,
either method works fine.

\begin{figure}
\includegraphics[width=4in]{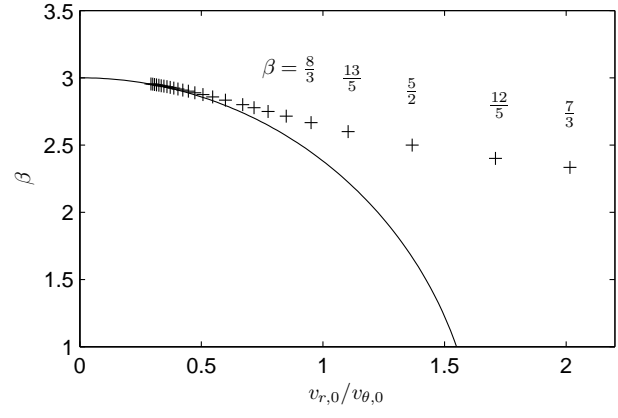}
  \caption{Relationship between $\beta$ and $v_{r,0}$ for $\lambda=6$ from Eq.~(\ref{eq:Browninvert6}) (solid line)
  and exact numerical calculation for closed orbits (crosses).
  The values of $\beta$ corresponding to the five largest amplitude closed orbits are also indicated.}
  \label{fig:lambda6beta}
\end{figure}

In principle,
orbits with any allowed value of $\beta$ can be found provided the initial guess for $v_{r,0}$ is accurate enough.
Even if Eq.~(\ref{eq:Browninvert}) does not supply a sufficiently accurate first guess,
the ``distance function'' \textit{versus} $v_{r,0}$ can easily be calculated and plotted,
and a better first guess obtained.
For $\lambda=6$,
several closed orbits were found using this method,
and the values of $\beta$ and $v_{r,0}$ for each orbit are shown in Fig.~\ref{fig:lambda6beta}.
The small amplitude relationship,
Eq.~(\ref{eq:Browninvert6}), is also shown,
and it can be seen that the two deviate when the radial amplitude becomes large.

What do these large amplitude orbits look like?
Besides the orbits with large values of $q$, which are close to circular,
the crosses in Fig.~\ref{fig:lambda6beta} indicate a few orbits with small values of $q$
(of course with $\beta$ still less than 3).
The orbit with the smallest value of $q$ is $\beta = \frac{5}{2}$.
The initial radial velocity and amplitude predicted by Eqs.~(\ref{eq:Browninvert6}) and (\ref{eq:vr0andepsilon})
are $v_{r,0} / v_{\theta,0} = 0.908$,
and $\epsilon/r_0 = 0.303$.
Since this radial oscillation amplitude is large,
the small amplitude result in Eq.~(\ref{eq:Brownresult}) is not applicable.
A search of parameter space (using the secant method explained above) reveals that the necessary initial radial velocity is
$v_{r,0} / v_{\theta,0} = 1.36671$,
and this orbit is shown in Fig.~\ref{fig:lambda64}.
Even though it can be classified as spirograph-like,
because $\beta$ is a ratio of two small integers (and is greater than unity),
the orbit has the appearance of being more ``star''-like.
Other similar values of $\beta$,
for example the cross labeled $\frac{7}{3}$ in Fig.~\ref{fig:lambda6beta},
are consistent with orbits that also have a star-like appearance.
The radial displacement of the star-like orbit turns out \textit{not} to be centered on $r_0$,
which is to be expected from a large-amplitude, anharmonic oscillator.
A rough estimate from the numerical solution gives
$\epsilon/r_0 \approx 0.405$, which is significantly larger than that predicted by Eq.~(\ref{eq:vr0andepsilon}).


\subsection{Large amplitude orbits and non-crossing orbits}

For a given value of $\lambda$, what is the range of possible values of $\beta$?
We have seen that for $\lambda=6$,
$\beta$ must remain less than three.
For other values of $\lambda$, $\beta$ is also restricted,
and this restriction is determined by Eq.~(\ref{eq:Brownresultv}),
which shows that $\beta$ must be either greater than or less than $\sqrt{\lambda +3}$
depending on the sign of the coefficient of $v_{r,0}^2$.
For $\lambda = 6$,
the coefficient is negative, $-(\lambda-1)(\lambda+2) = -40 < 0$,
which means that $\beta \le 3$, as we have already discovered.
This result divides the parameter space into three regimes,
and the boundaries between these regimes are just the two special cases of Bertrand's theorem:
\begin{equation}
  \label{eq:betalimits}
  \begin{array}{lrlll}
    \textrm{regime I  }   & \lambda < -2 \;\;\;\;\;    &&& \beta \le \sqrt{\lambda +3} \\
    \textrm{regime II }   & -2 < \lambda < 1 &&&  \beta \ge \sqrt{\lambda +3} \\
    \textrm{regime III  } & 1 < \lambda      &&& \beta \le \sqrt{\lambda +3}
  \end{array}
\end{equation}

\begin{figure}
\includegraphics[width=4in]{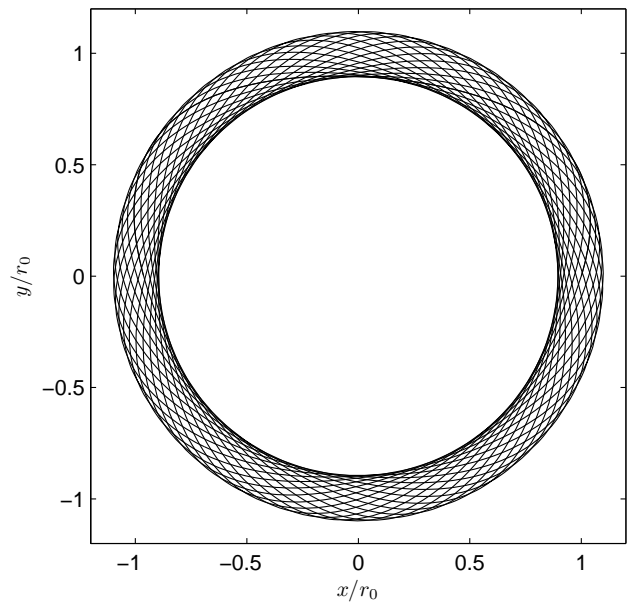}
  \caption{Same as Fig.~\ref{fig:lambda62}(a), except that $v_{r,0} / v_{\theta,0} \equiv 0.30748$.
  The orbit is closed, with $\beta = \frac{59}{20}$.
  The value that the small amplitude approximation, Eq.~(\ref{eq:Browninvert6}),
  predicts for the radial impulse is $v_{r,0} / v_{\theta,0} \approx 0.29875$.
  However, the radial amplitude is too large, $\epsilon/r_0 \approx 0.1$,
  for this prediction to be exact.
  A numerical root finding search was used to obtain the correct value.}
  \label{fig:lambda63}
\end{figure}

\begin{figure}
\includegraphics[width=4in]{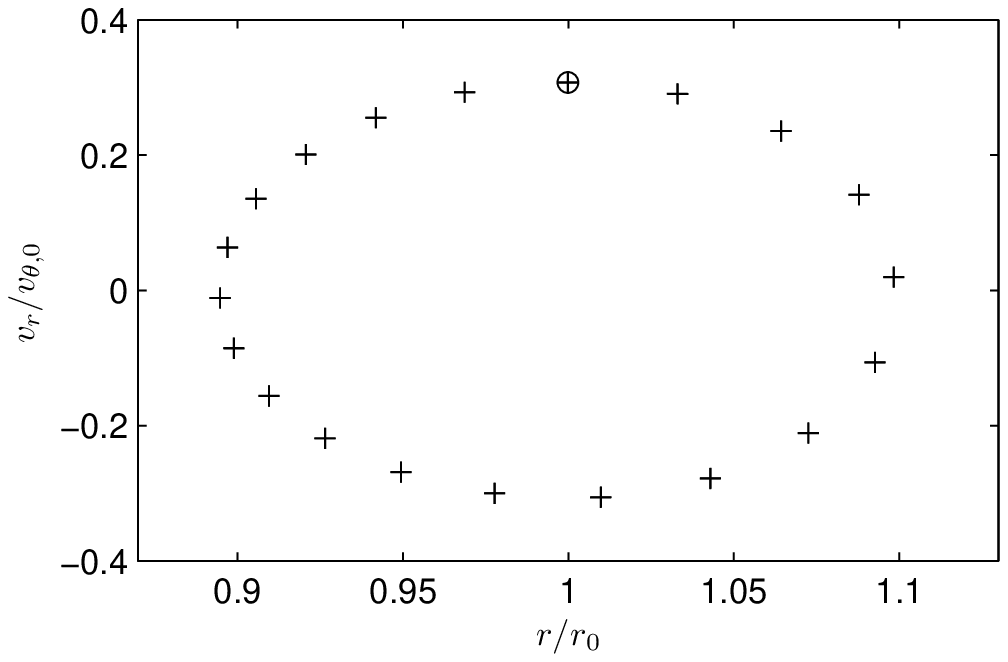}
  \caption{Poincare surface-of-section for the parameters in Fig.~\ref{fig:lambda63}.
  The trajectory's location in $(r,v_r)$ phase space is plotted each time it crosses the positive $x$-axis.
  Forty revolutions about the center are shown,
  so that each cross is really two crosses,
  depicting subsequent passings.
  The fact that the crossing locations are identical means that the orbit is closed.}
  \label{fig:lambda6phase}
\end{figure}

\begin{figure}
\includegraphics[width=4in]{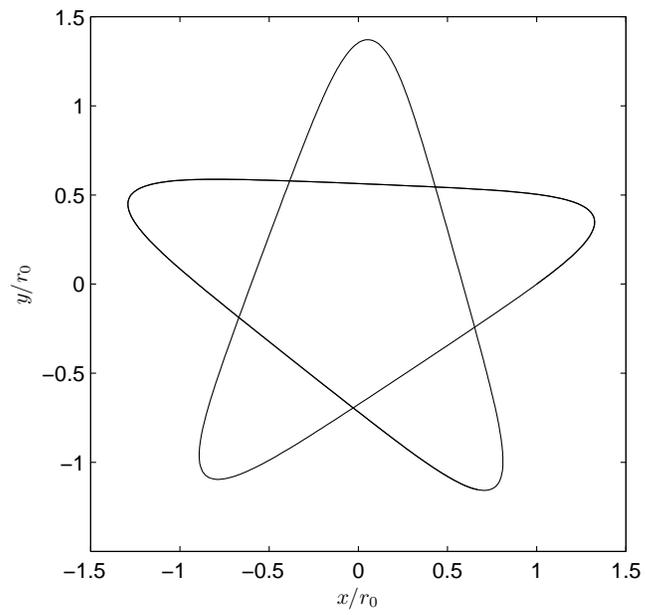}
  \caption{Same as Fig.~\ref{fig:lambda63}, except that $v_{r,0} / v_{\theta,0} \equiv 1.36671$.
  The orbit is closed, with $\beta = \frac{5}{2}$,
  and the radial amplitude is approximately $\epsilon/r_0 \approx 0.405$.}
  \label{fig:lambda64}
\end{figure}

\begin{figure}
\includegraphics[width=4in]{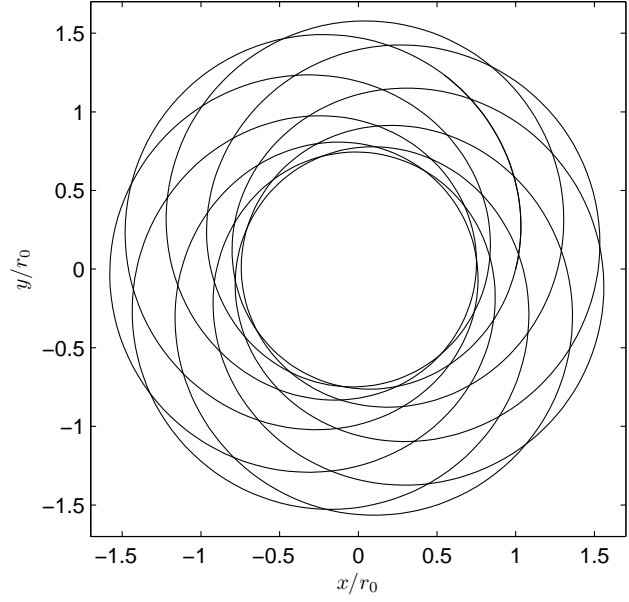}
  \caption{Same as Fig.~\ref{fig:lambda63}, except that $\lambda = -2.5$, $v_{r,0} / v_{\theta,0} \equiv 0.25034$.
  The orbit is closed, with $\beta = \frac{7}{10}$.
  Again, the anharmonicity of the potential means that the radial amplitude is not symmetric about $r_0$.
  The numerical result gives $r_{min} \approx 0.743 r_0$ and $r_{max} \approx 1.581 r_0$.}
  \label{fig:lambda2510}
\end{figure}

\begin{figure}
\includegraphics[width=4in]{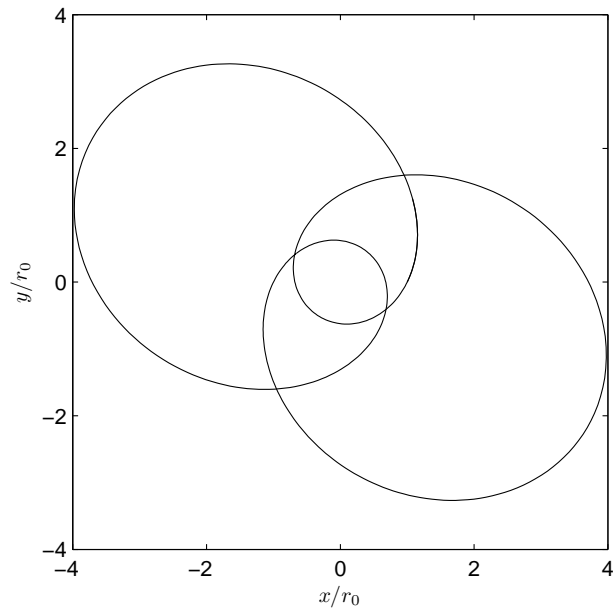}
  \caption{Same as Fig.~\ref{fig:lambda2510}, except that $v_{r,0} / v_{\theta,0} \equiv 0.48733$.
  The orbit is closed, with $\beta = \frac{2}{3}$.}
  \label{fig:lambda253}
\end{figure}

Regime I is actually restricted to $-3 < \lambda < -2$ because
we are only interested in bounded orbits.
The character of these orbits can be different from those we have already studied,
because $\beta$ will always be less than one.
In fact,
as we have defined it,
$\beta$ must be positive definite, so for regime I it must lie in the range
$0<\beta \le \sqrt{\lambda+3}$.
If $\lambda = -2.5$, for example,
$\beta \le 1/\sqrt{2} \approx 0.707$.
In this regime,
when $\beta$ is a ratio of two fairly large integers,
the orbits are similar to the orbit in Fig.~\ref{fig:lambda63}.
To see this,
the case of $\lambda=-2.5$ and $\beta=\frac{7}{10}$ is shown in Fig.~\ref{fig:lambda2510}.
The only difference in character between the two orbits is that in Fig.~\ref{fig:lambda2510} the number of orbits
is greater than the number of ``furthest approaches'' where $r=r_{max}$,
rather than vice-versa.
On the other hand,
regime I allows a new type of orbit because
when $\beta$ is the ratio of two small integers,
the nature of the trajectory radically changes.
Again,
for $\lambda=-2.5$,
the closed orbit where $\beta=\frac{2}{3}$ is shown in Fig.~\ref{fig:lambda253}.
Because only two values of $r_{min}$ (and $r_{max}$)
can occur during the course of three orbits,
the trajectory looks very different from Fig.~\ref{fig:lambda63}.
In fact, this orbit appears more ``loop''-like than spirograph-like.
This character comes from the fact that $\beta$ is the ratio of two small integers and is \textit{less} than one.
Regime I is the only case where $\beta$ can be less than unity.

\begin{figure}
\includegraphics[width=4in]{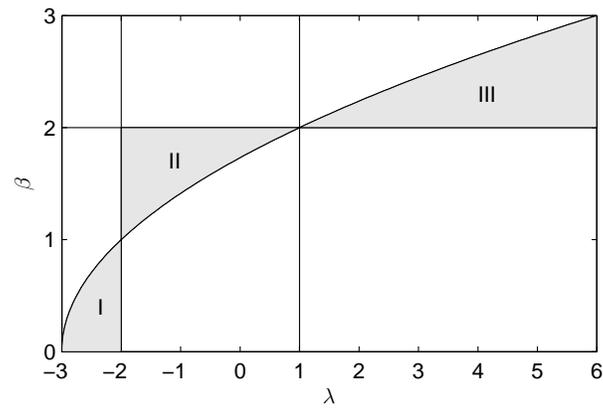}
  \caption{Allowed values of $\beta$ for the three regimes.
  The curve is the linear relationship between $\beta$ and $\lambda$ for nearly circular orbits, given by
  Eq.~(\ref{eq:betalambdaplus3}).
  The shaded regions indicate allowed values of $\beta$ for large radial oscillations in the three regimes,
  whose limits are given by Eq.~(\ref{eq:betalimits}) and in the subsequent text.}
  \label{fig:betalambda}
\end{figure}

In addition to non-power law forces,
Struck\cite{Struck06} focused on power-laws in regime II,
because these describe galactic potentials well.
He showed that in addition to the criterion in Eq.~(\ref{eq:betalimits}),
$\beta$ was restricted to $\beta <2$.
The global analysis of Bertrand\cite{Bertrand} also reveals this fact,
and in particular shows that $\beta \rightarrow 2$ in the limit that $r_{max} \rightarrow \infty$.
(For the numerical solutions in this study, this limit corresponds to $v_{r,0}/v_{\theta,0} \rightarrow \infty$.)
This limit also explains why the exact numerical solutions in Fig.~\ref{fig:lambda6beta} are all in the range $2 < \beta < 3$.
The allowed values of $\beta$ for all three regimes are shown in Fig.~\ref{fig:betalambda}.
Because of the restriction $1 < \beta < 2$ in regime II,
the types of orbits have the same character as Fig.~\ref{fig:lambda63}.
That is, they are of the spirograph type,
and except for $\lambda = -2$ and $\lambda = 1$,
they cannot be non-crossing.

\begin{figure}
\includegraphics[width=4in]{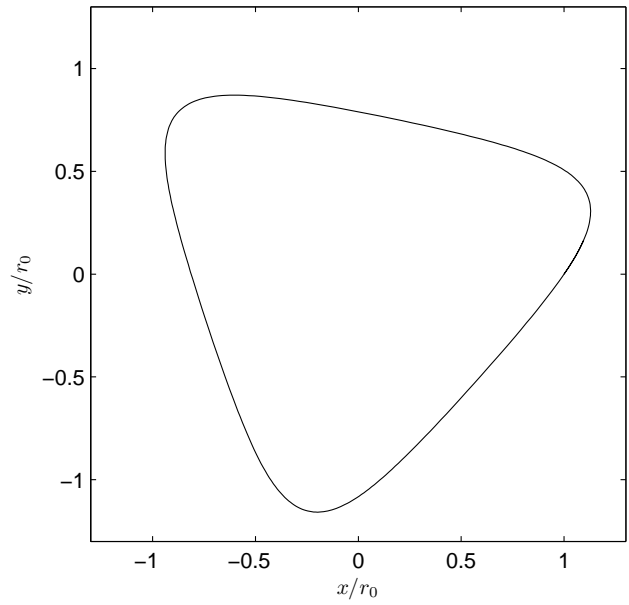}
  \caption{Same as Fig.~\ref{fig:lambda63}, except that $\lambda = 8$ and $v_{r,0} / v_{\theta,0} \equiv 0.68492$.
  The orbit is closed, with $\beta = \frac{3}{1}$.}
  \label{fig:lambda81}
\end{figure}

\begin{figure}
\includegraphics[width=4in]{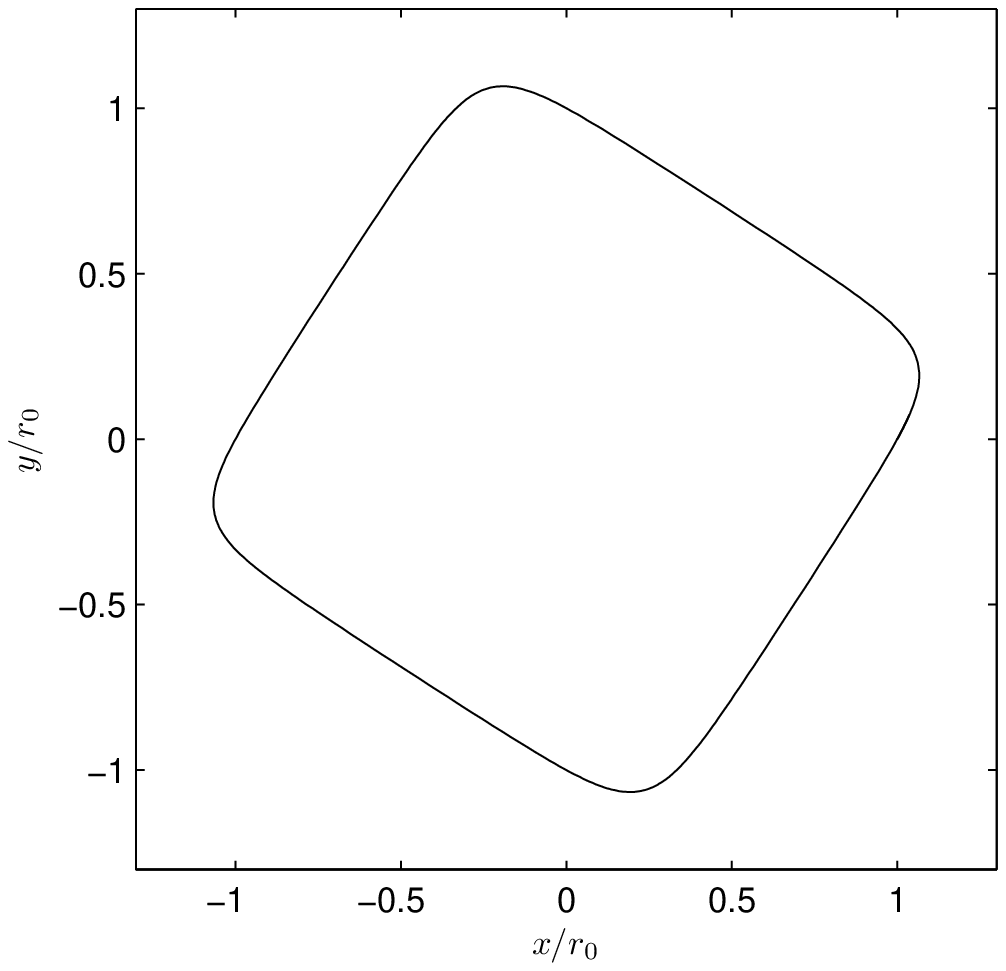}
  \caption{Same as Fig.~\ref{fig:lambda81}, except that $\lambda = 20$ and $v_{r,0} / v_{\theta,0} \equiv 0.54372$.
  The orbit is closed, with $\beta = \frac{4}{1}$.}
  \label{fig:lambda201}
\end{figure}

The final type of orbit with an interesting character occurs only in regime III.
These are characterized by $q=1$,
which means that the orbit is non-crossing.
The case of $\lambda=6$, studied above, does not admit a large-amplitude, non-crossing orbit,
since $\beta = \frac{3}{1}$ is restricted to the nearly circular case,
and $\beta = \frac{2}{1}$ is not accessible with a finite value of $v_{r,0}$.
However,
if $\sqrt{\lambda+3}>3$,
then a finite amplitude orbit can be consistent with $\beta = \frac{3}{1}$,
resulting in a closed, non-crossing, ``triangular''-shaped orbit.
This is shown in Fig.~\ref{fig:lambda81} for the force law parameter $\lambda=8$.
Any value of $\lambda$ greater than 6 will, of course,
admit a triangular orbit if $v_{r,0}$ has the proper value.
``Square''-shaped orbits can also occur when $\sqrt{\lambda+3}>4$,
i.e., $\lambda > 13$,
and one is shown in Fig.~\ref{fig:lambda201} where $\lambda = 20$.
Higher order ``polygonal'' orbits are also possible,
but they require increasingly larger minimum values of $\lambda$.

\section{Conclusion}
\label{sec:conclusion}

Closed orbit trajectories of several different types have been found for central forces
that are of a power-law type.
Besides the well-known elliptical orbits that arise from Coulomb's law ($\sim r^{-2}$) and Hooke's law ($\sim r$),
we have shown that closed orbits exist for all power law central forces,
$F_r = -kr^\lambda$, when $\lambda > -3$.
Over the largest part of parameter space,
the closed orbits are spirograph-like,
with many self crossings before they return to their original location.
However,
when $\beta$ is a ratio of two small integers,
then the orbits become more ``star''-like (Fig.~\ref{fig:lambda64})
or ``loop''-like (Fig.~\ref{fig:lambda253}).
Finally,
non-crossing orbits, when $\beta$ is an integer, occur for large values of $\lambda$,
and can be triangular, square, or polygonal.

\section*{Acknowledgements}

The authors would like to thank J. M. Hughes for useful discussions.

\newpage

\newpage

\newpage

\newpage

\newpage

\newpage

\newpage

\newpage

\newpage

\newpage

\newpage

\begin{thebibliography}{}
\bibitem{ref:smallperturb}
This result is derived in many textbooks,
notably in Ref.\ 2, \S 3-6.
\bibitem{energy}
The result in Eq.~(\ref{eq:vr0andepsilon}) can also be obtained from
an application of conservation of energy.
From a stable circular orbit, a radial impulse increases the kinetic energy,
which becomes potential energy at the maximum radial distance, $r_0 + \epsilon$.
\bibitem{lambda6}
The parameter $\lambda=6$ was chosen simply because it results in a small integer value for $\beta$.
\end{thebibliography}
\end{document}